\input phyzzx
\hsize=417pt 
\sequentialequations
\Pubnum={ EDO-EP-13}
\date={ \hfill May 1997}
\titlepage
\vskip 32pt
\title{ Wave Function of Evaporating Black Holes }
\vskip 12pt
\author{Ichiro Oda \footnote\dag {E-mail address: 
ioda@edogawa-u.ac.jp}}
\vskip 12pt
\address{ Edogawa University,                                
          474 Komaki, Nagareyama City,                        
          Chiba 270-01, JAPAN     }                          
%
%
%
%
%
\abstract{ We study some quantum mechanical aspects of dynamical black 
holes where the Vaidya metric is used as a model representing 
evaporating black holes. It is shown that in this model the 
Wheeler-DeWitt equation is solvable in whole region of spacetime, 
provided that one considers the ingoing (or outgoing) Vaidya metric and 
selects a suitable coordinate frame. This  
wave function has curious features in that near the curvature singularity 
it oscillates violently owing to large quantum effects while in the other 
regions the wave function exhibits a rather benign and completely regular 
behavior. The general formula concerning the black hole radiation, which 
reduces to the Hawking's semiclassical result when $r = 2M$ is chosen, is 
derived by means of purely quantum mechanical approach. The present 
formulation can be applied essentially to any system with a spherically 
symmetric black hole in an arbitrary spacetime dimension.
}
\endpage
%
%
%

\def\sp(#1){\noalign{\vskip #1pt}}

%
%
%
%
%
\topskip 30pt
\par

Hawking's discovery that quantum black holes can evaporate by emitting 
thermal radiation [1] has stimulated not only active researches on 
quantum field theory in curved spacetime and a theory of quantum gravity 
but also a great deal of speculation about a grand synthesis of general 
relativity, thermodynamics, and quantum theory [2].  
Unfortunately, however, in spite of much impressive effort it might be 
fair to say that in four dimensions little progress has been made in 
generalizing Hawking's ``semiclassical'' analysis to mathematically 
self-consistent, quantum mechanically correct approach. 

Recently, we have developed a new approach which is purely quantum 
mechanical in that not only the matter fields but also the gravitational 
field are treated as quantum fields, in order to understand the black 
hole radiation [3, 4], and subsequently applied fruitfully to various 
problems relevant to quantum black holes, for example, the mass inflation 
in the Reissner-Nordstrom charged black hole [5], the three dimensional 
de Sitter black hole [6], the weak cosmic censorship [7], and the two 
dimensional dilaton gravity [8]. The key observation in this approach is 
that we can carry out the canonical quantization for a spherically 
symmetric system with the Vaidya metric representing the dynamical black 
holes if we limit our consideration to only the region near the apparent 
horizon and at the same time choose an appropriate coordinate frame. 

However, this approach gives us some frustrations. This is because within 
the approach it is impossible to gain useful informations about quantum 
structure of spacetime in whole region through the wave function 
and it is also quite unclear what approximation method considering the 
specific region of spacetime corresponds to.  
 
The main motivation in this paper is to remove these disadvantages and put 
our approach  on a more sound foundation. As a bonus, we 
will see that the wave function obtained in this way shows very 
interesting, physically reasonable features such that it fluctuates 
violently near the curvature singularity while it behaves mildly in the 
other regions. In addition, we can derive the more general formula with 
respect to the mass loss rate of an evaporating black hole compared to 
the semiclassical approach.

The action which we consider in this paper is of form
$$ \eqalign{ \sp(2.0)
S = \int  d^4 x \sqrt{-^{(4)}g} \ \bigl( \ {1 \over 16 \pi G} {}^{(4)}R - 
{1 \over  8 \pi} {}^{(4)}g^{\mu\nu} \partial_{\mu} \Phi \partial_{\nu} 
\Phi \  \bigr),
\cr
\sp(3.0)} \eqno(1)$$
where $\Phi$ is a real scalar field. To show the four dimensional  
character explicitly we put the suffix $(4)$ in front of the metric 
tensor and the curvature scalar. We follow the conventions adopted in the 
MTW textbook [9] and use the natural units $G = \hbar = c = 1$. The Greek 
indices $\mu, \nu, ...$ take the values 0, 1, 2, and 3, while the Latin 
indices $a, b, ...$ take the values 0 and 1. Of course, the inclusion of 
other matter fields, the cosmological constants and the surface terms in 
this formalism is straightforward even if we confine ourselves to the 
action (1) for the sake of simplicity. 

After a general spherically symmetric reduction [10]
$$ \eqalign{ \sp(2.0)
ds^2 &= {}^{(4)}g_{\mu\nu} dx^{\mu} dx^{\nu},
\cr
     &= g_{ab}(x^c) dx^a dx^b + \phi(x^c)^2 ( d\theta^2 + \sin^2\theta 
     d\varphi^2 ),  
\cr
\sp(3.0)} \eqno(2)$$
the action (1) can be rewritten as
$$ \eqalign { \sp(2.0)
S &= {1 \over 2} \int  d^2 x \sqrt{-g} \ \bigl( 1 + g^{ab} \partial_a \phi 
\partial_b \phi +  {1 \over 2} R \phi^2 \bigr)
\cr
&\qquad- {1 \over 2} \int  d^2 x \sqrt{-g} \ \phi^2 g^{ab} \partial_a 
\Phi \partial_b \Phi.
\cr
\sp(3.0)} \eqno(3)$$
 
Moreover, introducing the ADM parametrization 
$$ \eqalign{ \sp(2.0)
g_{ab} = \left(\matrix{ { - \alpha^2 + {\beta^2 \over \gamma}} & \beta \cr
              \beta & \gamma \cr} \right),
\cr
\sp(3.0)} \eqno(4)$$
and $n^a$ which is unit vector normal to the foliations $x^0 = const$  
$$ \eqalign{ \sp(2.0)
n^a = ( {1 \over \alpha}, \ - {\beta \over {\alpha \gamma}}),
\cr
\sp(3.0)} \eqno(5)$$
the action (3) becomes
$$ \eqalign{ \sp(2.0)
S &= \int d^2x \ \bigl[ \ {1 \over 2} \alpha \sqrt{\gamma} \ \bigl\{ 1 - 
(n^a \partial_a \phi)^2 + {1 \over \gamma} (\phi^\prime)^2 - K n^a 
\partial_a (\phi^2)  
\cr
&\qquad+ {\alpha^\prime \over {\alpha\gamma}} \partial_1 (\phi^2) \bigr\} 
+ {1 \over 2} \alpha \sqrt{\gamma} \ \phi^2 \bigl\{ ( n^a \partial_a 
\Phi )^2 - {1 \over \gamma} ( \Phi^\prime )^2 \bigr\} \ \bigr],
\cr
\sp(3.0)} \eqno(6)$$
where $K$ is the trace of the extrinsic curvature defined as ${1 \over 
\sqrt{-g}} \partial_a  ( \sqrt{-g} \ n^a )$ and ${\partial \over {\partial 
x^0}} = \partial_0$  and ${\partial \over {\partial x^1}} = \partial_1$ 
are also denoted by an overdot and a prime, respectively. 

{}From (6), the canonical conjugate momenta can be read off 
$$ \eqalign{ \sp(2.0)
p_{\Phi} &= \sqrt{\gamma} \ \phi^2 n^a \partial_a \Phi,
\cr
p_{\phi} &= - \sqrt{\gamma} \ n^a \partial_a \phi - \sqrt{\gamma} \ K \phi,
\cr
p_{\gamma} &= - {1 \over 4 \sqrt \gamma} n^a \partial_a (\phi^2).
\cr
\sp(3.0)} \eqno(7)$$
Then the Hamiltonian can be found to be a linear combination of the 
Hamiltonian and the momentum constraints as follows:
$$ \eqalign{ \sp(2.0)
H = \int dx^1 \ ( \alpha H_0 + \beta H_1 ), 
\cr
\sp(3.0)} \eqno(8)$$
where the constraints are explicitly given by
$$ \eqalign{ \sp(2.0)
H_0 &= {1 \over 2 \sqrt{\gamma} \phi^2} p_{\Phi} ^2 - {\sqrt{\gamma} 
\over 2} - { (\phi^\prime)^2 \over 2 \sqrt{\gamma} } 
+ \partial_1 ( {\partial_1 (\phi^2) \over 2 \sqrt{\gamma}}) 
\cr
&\qquad+ {\phi^2 \over 2 \sqrt{\gamma}} ( \Phi^\prime )^2 - {2 
\sqrt{\gamma} \over \phi} p_\phi p_\gamma + {2 \gamma \sqrt{\gamma} \over 
\phi^2} p_\gamma ^2 , 
\cr
\sp(3.0)} \eqno(9)$$
$$ \eqalign{ \sp(2.0)
H_1 = {1 \over \gamma} \ p_\Phi \Phi^\prime + {1 \over \gamma} p_\phi 
\phi^\prime - 2  p_\gamma ^\prime - {1 \over \gamma} p_\gamma 
\gamma^\prime. 
\cr
\sp(3.0)} \eqno(10)$$

The canonical formalism explained so far gives a basis to construct a 
quantum theory of a system with a spherically symmetric black hole in the 
below. As a first crucial step toward the canonical quantization, we shall 
introduce the two dimensional coordinate $x^a$ by $\footnote 1 {This 
choice of the coordinate system is very critical to reach the desired 
result. For instance, instead of (11) if we choose $x^a  = (v, r)$ which 
reduces to the Eddington-Finkelstein coordinate in vacuum, later 
analysis would reveal that the wave function is independent of the matter 
field and static.}$ 
$$ \eqalign{ \sp(2.0)
x^a = (x^0, x^1) = (v - r, \  r),
\cr
\sp(3.0)} \eqno(11)$$
where $v$ is the advanced time coordinate. Next let us fix the gauge 
symmetries by the gauge conditions 
$$ \eqalign{ \sp(2.0)
g_{ab} &= \left(\matrix{ { - \alpha^2 + {\beta^2 \over \gamma}} & \beta \cr
              \beta & \gamma \cr} \right),
\cr
 &= \left(\matrix{ -(1 - {2M \over r})   &  {2M \over r} \cr
              {2M \over r}  & 1 + {2M \over r} } \right),
\cr
\sp(3.0)} \eqno(12)$$
where the black hole mass $M$ is in general the function of the two 
dimensional coordinate $x^a$. From (11) and (12) the two dimensional line 
element takes a form
$$ \eqalign{ \sp(2.0)
ds^2 &= g_{ab} dx^a dx^b,
\cr
     &= -(1 - {2M \over r}) dv^2 + 2 dv dr.
\cr
\sp(3.0)} \eqno(13)$$

In previous works [3-8], the model which we have described till now and 
its variants have been widely used to perform the canonical quantization 
of dynamical black holes near the apparent horizon. The main new 
achievement in this article is to accomplish the quantization not only 
near the apparent horizon but over the whole spacetime region. This will 
enable the wave function to be explored in considerable detail. To do so, 
since we are interested in the ingoing Vaidya metric [11], we impose the 
reasonable assumption on the dynamical fields 
$$ \eqalign{ \sp(2.0)
\Phi = \Phi(v), \ M = M(v), \ \phi = r, 
\cr
\sp(3.0)} \eqno(14)$$
which is obviously consistent with the fields equations stemming from 
(1). Under the assumption (14), the canonical conjugate momenta (7) and 
the constraints (9) and (10) reduce to 
$$ \eqalign{ \sp(2.0)
p_{\Phi} &=  \phi^2 \ \partial_v \Phi,
\cr
p_{\phi} &= {1 \over \gamma} \partial_v M + {2 M^2 \over r^2 \gamma},
\cr
p_{\gamma} &= {M \over \gamma},
\cr
\sp(3.0)} \eqno(15)$$
$$ \eqalign{ \sp(2.0)
\sqrt{\gamma} H_0 &= \gamma  H_1,
\cr
             &= {1 \over \phi^2} \ p_{\Phi}^2  - \gamma \ p_\phi + {2 
             M^2  \over r^2}, 
\cr
\sp(3.0)} \eqno(16)$$
where $\gamma = 1 + {2 M \over r}$ from (12). Here some comments are in 
order. First of all, the reason why the two constraints are proportional 
to each other in (16) is that there remains only one residual symmetry 
which generates the translation $v \rightarrow v + \varepsilon$ owing to 
the assumption (14)  
whose situation bears some resemblance to the case of the lightcone 
string field theory [12]. This corrects the wrong reasoning stated in 
previous our work [6]. Secondly, the dynamical degrees of freedom $\gamma$ 
corresponding to ``graviton'' are frozen because $p_{\gamma}$ in (15) does 
not have the term including a differentiation with respect to $v$. This 
is nothing but a manifestation of the Birkoff's theorem [9]. Finally, in 
(16), $p_{\phi}$ appears in the linear form so that this equation has the 
same structure as the Schrodinger equation if we regard a suitable 
function of $\phi$ as the time in the superspace.

Substituting $p_{\Phi} = - i {\partial \over \partial \Phi}$ and 
$p_{\phi} = - i {\partial \over \partial \phi}$ into the constraint (16) 
yields the Wheeler-DeWitt equation.
$$ \eqalign{ \sp(2.0)
( - {1 \over \phi^2} {\partial^2 \over \partial \Phi^2} + i \gamma  
{\partial \over \partial \phi} + {2 M^2  \over r^2} ) \Psi = 0.  
\cr
\sp(3.0)} \eqno(17)$$
It is straightforward to obtain a special solution of (17) by the method 
of separation of variables. The result is 
$$ \eqalign{ \sp(2.0)
\Psi = (  B e^{\sqrt{A} \Phi(v)} + C e^{-\sqrt{A} \Phi(v)} ) \ e^{ i {{ A 
- 2 M^2 } \over 2 M} \log \gamma },
\cr
\sp(3.0)} \eqno(18)$$
where $A$, $B$, and $C$ are integration constants. Let us examine the 
physical implications of this wave function in detail.  At the spatial 
infinity $r  
\rightarrow  \infty$, $\gamma \rightarrow 1$ so that $\Psi$ consists of 
only the ingoing matter field in the asymptotically flat spacetime as 
expected. And at the apparent horizon $r = 2 M(v)$, $\gamma = 2$ so 
that $\Psi$ is composed of the part of the ingoing matter field plus the 
part affected by the gravitational field there. Finally in the vicinity 
of the curvature singularity $r \rightarrow  0$, $\gamma \rightarrow 
\infty$ thus $\Psi$ oscillates violently. It is important to remember 
that the field $\gamma$ is exactly the dynamical degrees of the freedom 
of ``graviton''. Although classically these degrees of the freedom are 
killed owing to a spherically symmetric ansatz, it is of interest that 
they reappear and play an important role in quantum theory through the wave 
function. In other words, we can regard that strong quantum effects 
associated with the gravitational degrees of freedom cause the wave 
function to be singular at the singularity.  Note that $\Psi$ is 
completely regular over the whole region of spacetime except at the 
curvature singularity. Also notice that in deriving the Wheeler-DeWitt 
equation (17) and its solution (18) we have never appealed to any 
approximation method, thus they are essentially in the nonperturbative 
regime.

Now under a general definition of the expectation value, the change rate 
of the mass function of a black hole can be evaluated by means of (16) 
and (18) 
$$ \eqalign{ \sp(2.0)
< \partial_v M > = - {A \over r^2}.
\cr
\sp(3.0)} \eqno(19)$$
If we set $r$ to be equal to the black hole radius $2M$, (19) 
reproduces the Hawking's semiclassical result up to a numerical constant [1]. 
But (19) is derived from the purely quantum mechanical arguments and  
provides the more general result than the semiclassical one although I 
have no idea how to determine the constant $A$ within the present 
framework. In order to fix the numerical coefficient in this 
formula, maybe we would need to impose the boundary conditions 
on the wave function. Furthermore, this formula strongly suggests that 
the particles  
emitted by an evaporating black hole can be traced all the way back to 
not the surface of the horizon but the singularity. 
Incidentally, when we consider the outgoing Vaidya metric 
$$ \eqalign{ \sp(2.0)
ds^2 = -(1 - {2M(u) \over r}) du^2 - 2 du dr
\cr
\sp(3.0)} \eqno(20)$$
instead of the ingoing one (13), we can repeat the previous arguments and 
it then turns out that we can reach similar results to the case of the 
ingoing Vaidya metric except a minor change where the coordinate $v$ must 
be replaced by $u$. One weakness in the present model is the lack of 
outgoing radiation escaping to the future null infinity. It is valuable 
to mention that the loss of the black hole mass in (19) is controlled by 
sending in negative energy flux from the past null infinity, rather than 
by having positive energy flux radiated to the future null infinity. In 
future, it would be interesting to construct a model which takes account 
of both the energy fluxes at the same time.

It is of interest to inquire whether the wave function has also
a singular behavior near the the curvature singularity $r = 0$ when 
we use a different kind of hypersurfaces, for example, $x^1 = r = const$ 
which would be more appropriate inside the apparent horizon. Using the 
canonical formalism constructed previously [4] 
and similar thoughts to the above, it is straightforward to derive the 
Wheeler-DeWitt equation 
$$ \eqalign{ \sp(2.0)
\bigl[  - {1 \over \phi^2} {\partial^2 \over \partial \Phi^2} - i \gamma  
{\partial \over \partial \phi} - \gamma \ (1 - { M  \over r } ) \bigr] \Psi 
= 0,   
\cr
\sp(3.0)} \eqno(21)$$
where this time $\gamma$ is given by $\gamma = - ( 1 - {2M \over r} ) > 
0$ inside the apparent horizon. The wave function satisfying (21) is 
$$ \eqalign{ \sp(2.0)
\Psi = (  B e^{\sqrt{A} \Phi(v)} + C e^{-\sqrt{A} \Phi(v)} ) \ e^{ i ( r 
- M \log r ) - { i A \over 2 M } \log \gamma }.
\cr
\sp(3.0)} \eqno(22)$$
It is obvious that near the singularity the wave function (22) 
oscillates violently owing to the factors $\log r$ as well as $\gamma$ 
which would be again rooted to strong quantum effects of the gravitational 
degrees of freedom. It is valuable to comment two points here. One is 
that we have  
previously obtained the wave function with similar features to this in a 
different context where the radial quantization was carried out for 
``static'' black hole models without matter field [13]. Their conformity 
in the behavior of the wave function suggests that the singular behavior 
of the wave function at the singularity has a connection with quantum 
effects of the gravitational field. Of course, the quantum effects 
associated with the matter field would make the singular behavior there
stronger. This point could be also understood from the fact that the change 
rate of the mass function (19) diverges at the singularity. The 
other is that we cannot connect smoothly the two wave 
functions (18) and (22) just on the apparent horizon. This is simply 
because we have used the different formulations.

In conclusion, we have investigated some quantum aspects of a dynamical 
black hole corresponding to the Schwarzschild geometry in four dimensions. 
However, the present analysis is easily generalized to all  
spherically symmetric black holes in any spacetime dimension. One 
intriguing question in future is how the present analysis takes over if 
we use the Kruskal coordinate which enables us to slice the spacetime by 
the hypersurfaces extending from the left spatial infinity to the right 
spatial infinity across the horizon and to find the wave function holding 
over the whole Kruskal extention of the Schwarzschild geometry. In this 
respect, Kuchar's work might be helpful [14].    

\vskip 32pt
\leftline{\bf Acknowledgments}
\centerline{ } %
\par
This work was supported in part by Grant-in-Aid for Scientific Research 
from Ministry of Education, Science and Culture No.09740212.

\vskip 32pt
\leftline{\bf References}
\centerline{ } %
\par
\item{[1]} S.W.Hawking, Comm. Math. Phys. {\bf 43}, 199 (1975) .

\item{[2]} N.D.Birrell and P.C.W.Davies, Quantum Fields in Curved Space 
(Cambridge University Press, Cambridge, 1982).

\item{[3]} A.Tomimatsu, Phys. Lett. {\bf B289}, 283 (1992) .

\item{[4]} A.Hosoya and I.Oda, Prog. Theor. Phys. {\bf 97}, 233 (1997).

\item{[5]} I.Oda, ``Mass Inflation in Quantum Gravity'', EDO-EP-8, 
gr-qc/9701058.

\item{[6]} I.Oda, ``Evaporation of Three Dimensional Black Hole in 
Quantum Gravity'', EDO-EP-9, gr-qc/9703055; ``Quantum Instability of 
Black Hole Singularity in Three Dimensions'', EDO-EP-10, gr-qc/9703056.

\item{[7]} I.Oda, ``Cosmic Censorship in Quantum Gravity'', 
gr-qc/9704021.   

\item{[8]} I.Oda, ``Quantum Gravity near Apparent Horizon and Two 
Dimensional Dilaton Gravity'', EDO-EP-12, gr-qc/9704084.

\item{[9]} C.W.Misner, K.S.Thorne, and J.A.Wheeler, Gravitation (Freeman, 
1973).  

\item{[10]} P.Hajicek, Phys. Rev. {\bf D30}, 1178 (1984); P.Thomi, B.Isaak  
and P.Hajicek, Phys.  Rev. {\bf D30}, 1168 (1984).

\item{[11]} P.C.Vaidya, Proc. Indian Acad. Sci. {\bf A33}, 264 (1951); 
P.Hajicek and W.Israel, Phys.  Lett. {\bf 80A}, 9 (1980); W.A.Hiscock, 
Phys. Rev. {\bf D23}, 2813, 2823 (1981).

\item{[12]} M.Kaku and K.Kikkawa, Phys. Rev. {\bf D10}, 1110, 1823 (1974).

\item{[13]} A.Hosoya and I.Oda, ``Black Hole Singularity and Generalized 
Quantum Affine Parameter'', TIT/HEP-334/COSMO-73, EDO-EP-5, gr-qc/9605069.

\item{[14]} K.V.Kuchar, Phys. Rev. {\bf D50}, 3961 (1994).

\endpage
%

%
\bye